\documentclass[reprint, superscriptaddress, amsmath, amssymb, aps, floatfix]{revtex4-2}

\usepackage[columnwise,running]{lineno}
\usepackage{orcidlink}
\usepackage{hyperref}
\usepackage{physics}
\usepackage{ragged2e}

\usepackage{graphicx}
\usepackage{dcolumn}
\usepackage{bm}

\usepackage[utf8]{inputenc}
\usepackage[T1]{fontenc}

\usepackage{braket}
\usepackage{siunitx}
\usepackage{xcolor}

\usepackage{booktabs}
\usepackage{multirow}

\usepackage[normalem]{ulem}

\def\FF{\mathcal{F}}                                    

\def\ddt{\dd t}
\def\ddW{\dd W}
\def\ddD{\dd D}

\def\dds{\dd s}

\newcommand{\sev}[1]{\langle #1 \rangle}

\begin{document}

\title{Quantum Fokker-Planck Master Equation with general signal filtering}
\date{\today}

\author{Guilherme De Sousa\ \orcidlink{0000-0002-8529-5439}}
\email{guilherme2.desousa@gmail.com}
\affiliation{Department of Physics$,$ University of Maryland$,$ College Park$,$ MD 20742$,$ USA}
\affiliation{Instituto de Física de São Carlos$,$ Universidade de São Paulo$,$ 13560-970 São Carlos$,$ SP$,$ Brasil}

\begin{abstract}
In this paper, we derive a general master equation for continuous feedback based on arbitrary linear signals.
This result is an extension of the Quantum Fokker-Planck Master Equation derived by Annby-Andersson \emph{et al} (Phys. Rev. Lett. \textbf{129}, 050401) to the case where the experiment has a general filtering structure.
The filtering operation can include delayed information, memory effects, and non-Markovian signal processing.
Using this general master equation, we derive analytical results for the ground state cooling of a quantum harmonic oscillator.
We compared our results with those derived by De Sousa \emph{et al} (Phys. Rev. E \textbf{111}, 014152).
Our framework aims to capture more realistic situations, to allow experiments to be better modeled, and to study non-Markovian effects in the spectral properties of the measurement signal.
\end{abstract}

\maketitle
\section{Introduction}
The ability to measure and control quantum systems in real time has advanced dramatically over the last decades.
Experimental platforms ranging from cold atoms \cite{Metcalf2003}, solid-state devices \cite{Gustavsson2006,Barker2019}, and quantum optical setups \cite{Hadfield2009,Charaev2023} now routinely achieve levels of precision where feedback and signal processing are essential components of quantum experiments.
These developments have fueled a growing interest in quantum feedback control and continuous measurement as important tools in quantum information science and quantum thermodynamics \cite{Lloyd2000,Geremia2004,Wiseman2009,Sayrin2011,Zhang2017}.

At the theoretical level, continuous quantum measurement 
is often modeled using quantum unraveling, which can be done through quantum jumps~\cite{Pettruccione2007,Binder2019}, quantum state diffusion~\cite{Diosi1988,Belavkin1989,Belavkin1989_2,Barchielli1991,Jacobs1998,Jacobs2006}, or both~\cite{AnnbyAndersson2024}.
In the context of quantum state diffusion, individual experimental trajectories are modeled as stochastic processes conditioned on measurement outcomes, allowing one to implement continuous feedback at the trajectory level~\cite{Wiseman1993}.
While this trajectory viewpoint has been extremely powerful, ensemble-based approaches can often provide complementary insights when optimizing protocols and understanding the asymptotic behavior of observables across many realizations of the experiment.



Ensemble-based approaches have been derived in the context of hybrid master equations~\cite{Layton2024,Diosi2014,Oppenheim2023,Diosi2023,Tilloy2024,Diosi1988,Oppenheim2022,Layton2025}, the Quantum Fokker-Planck Master Equation (QFPME)~\cite{AnnbyAndersson2022}, and feedback-resolved equations~\cite{Rosal2025} for generic  Markovian feedback.
The QFPME describes the joint evolution of the quantum system and the measurement outcome distribution~\cite{Kiely2025}, capturing the interplay between measurement backaction, information gain, and feedback.
Importantly, it provides a master-equation description that extends the Wiseman-Milburn formalism \cite{Wiseman1993} to nonlinear and delayed feedback.

The QFPME provides a powerful way to describe continuous feedback, but it has so far been limited to exponential low-pass response.
Real experiments, however, typically involve nontrivial signal processing.
Measurement outcomes are rarely fed directly back into the system. 
Instead, the signals undergo filtering due to detector bandwidth limitations, finite response times, or deliberately engineered spectral processing \cite{Grimsmo2015,Hirose2018,Vodenkova2024}.
In this work, we extend the QFPME framework to incorporate \emph{arbitrary linear filters} in the feedback loop.
This generalization makes the theory directly applicable to a wide range of experimental scenarios, including those where memory effects and non-Markovianity from the signal processing cannot be ignored.

Our goal is thus twofold: (i) to provide a general theoretical framework that allows experimentalists to faithfully model their feedback architectures within the QFPME formalism, and (ii) to demonstrate, through explicit examples, how ensemble-based master equations can yield optimization strategies that go beyond trajectory-level analyses.

The QFPME was derived for a continuously monitored system and feedback is applied based on the measurement outcome.
The continuous measurement~\cite{Jacobs2006} of a Hermitian operator $\hat{A}$ yields a signal that can be modeled by a Gaussian white noise
\begin{equation}
    \label{eq:signal_z}
    z(t) = \langle \hat{A} \rangle(t) + \frac{1}{\sqrt{4\lambda}} \frac{\dd W}{\dd t}.
\end{equation}
Here $z(t)$ denotes the measurement outcome; $\langle \hat{A} \rangle(t) = \Tr [\hat{A} \hat{\rho}(t)]$ is the expectation value of the operator; $\lambda$ is the measurement strength; and $\dd W$ is a stochastic Wiener increment.
Linear continuous feedback can be applied based on the signal in Eq.~\eqref{eq:signal_z}.
Once we average all possible trajectories, the system evolves under the formalism derived by Wiseman-Milburn \cite{Wiseman1993}.

The key ingredient introduced by the QFPME in Ref.~\cite{AnnbyAndersson2022} for non-linear feedback is to consider that real feedback schemes ``lag'' behind the signal $z(t)$.
The lagging signal $D(t)$ can be modeled using a low-pass filter:
\begin{align}
    \label{eq:signal_D}
    &D(t) = \gamma \int_{-\infty}^t \dds e^{-\gamma(t-s)} z(s),
\end{align}
where $\gamma$ is the low-pass frequency (detector bandwidth).
The QFPME is the equation that describes the time evolution of the joint distribution of the quantum state and the measurement outcome $D$.
The object of interest of the QFPME is a density matrix parametrized by the detector outcome
\begin{equation}
    \hat{\rho}(D,t) = \mathrm{E}_W[\hat{\rho}_c \delta (D-D(t))].
\end{equation}
Here, ${\rm E}_W[\cdot]$ represents the average over many realizations of the stochastic measurement outcome, and $\hat{\rho}_c$ describes the density matrix for a given measurement outcome trajectory $D{(t)}$. Throughout this work, we will denote by $D(t)$ a stochastic signal and by $D$ a particular value of that signal.

We use the same continuous measurement modeled by a Gaussian white noise, but consider a general filtering operation that can describe a larger class of linear systems.
The linearity of the filtering signal allows us to derive a general master equation for many classes of problems.
Using this generalized QFPME, we derive analytical results for the ground-state cooling of a quantum harmonic {oscillator}~\cite{DeSousa2025}, allowing us to identify parameter regimes where filtering enhances or suppresses cooling performance.
In particular, we identify nontrivial dependencies on the ratio between information collection and processing rates, and show how changing the filtering can lead to richer phase diagrams of cooling efficiency.
The results indicate that one can describe feedback systems using Markovian dynamics, even when the filters contain complicated frequency dependence.
This is due to an increase in the system's dimensionality, similar to a Markovian embedding \cite{Siegle2010,Campbell2018,Li2021}.

The paper is structured hierarchically, starting from the general formulation and then showing how it can be applied to particular cases of signal filtering.
We also apply it to a concrete example of a feedback system of a cooled harmonic oscillator.
The sections are organized as follows: Sec.~\ref{sec:qfpme_general} outlines the derivation of the main result of a QFPME for general signal processing.
Next, in Sec.~\ref{sec:modeling_filters}, we model three different filters using particular examples of the general model.
Section~\ref{sec:harmonic-oscillator} investigates the cooling of a harmonic oscillator using multiple low-pass filters and a band-pass filter. We discuss and conclude in Sec.~\ref{sec:discussion}.

\section{A QFPME with general filtering}\label{sec:qfpme_general}
Before we outline the derivation of the main results, let's introduce the setup.
Consider a quantum system that, in the absence of any measurement or feedback, evolves under the unitary dynamics described by a Hamiltonian $\hat{H}$.
The instantaneous state of the system is encoded in the density matrix $\hat{\rho}$.
At time $t$, we perform a weak measurement modeled by the action of a Gaussian Krauss operator~\cite{Jacobs2006,AnnbyAndersson2022}.
The evolution of the density matrix is described by
\begin{equation}
    \hat{\rho} \rightarrow e^{-i\delta t \hat{H}/\hbar} \frac{\hat{M}(z)\hat{\rho}\hat{M}^\dagger(z)}{\Tr[\hat{M}(z)\hat{\rho}\hat{M}^\dagger(z)]} e^{i\delta t \hat{H}/\hbar}.
\end{equation}
The duration of the measurement is given by $\delta t$, and is considered to be (infinitesimally) small.
The operator $\hat{M}(z)$ describes the Gaussian model for the measurement,
\begin{equation}
    \hat{M}(z) = \sqrt{\frac{2\lambda\delta t}{\pi}} e^{-\lambda \delta t (z-\hat{A})^2}.
\end{equation}
The operator $\hat{A}$ is the physical operator being measured, and $\lambda$ describes the measurement strength.
For simplicity, we consider $\hat{A} = \hat{A}^\dagger$ to be Hermitian.

\begin{figure}
    \centering
    \includegraphics[width=1\linewidth]{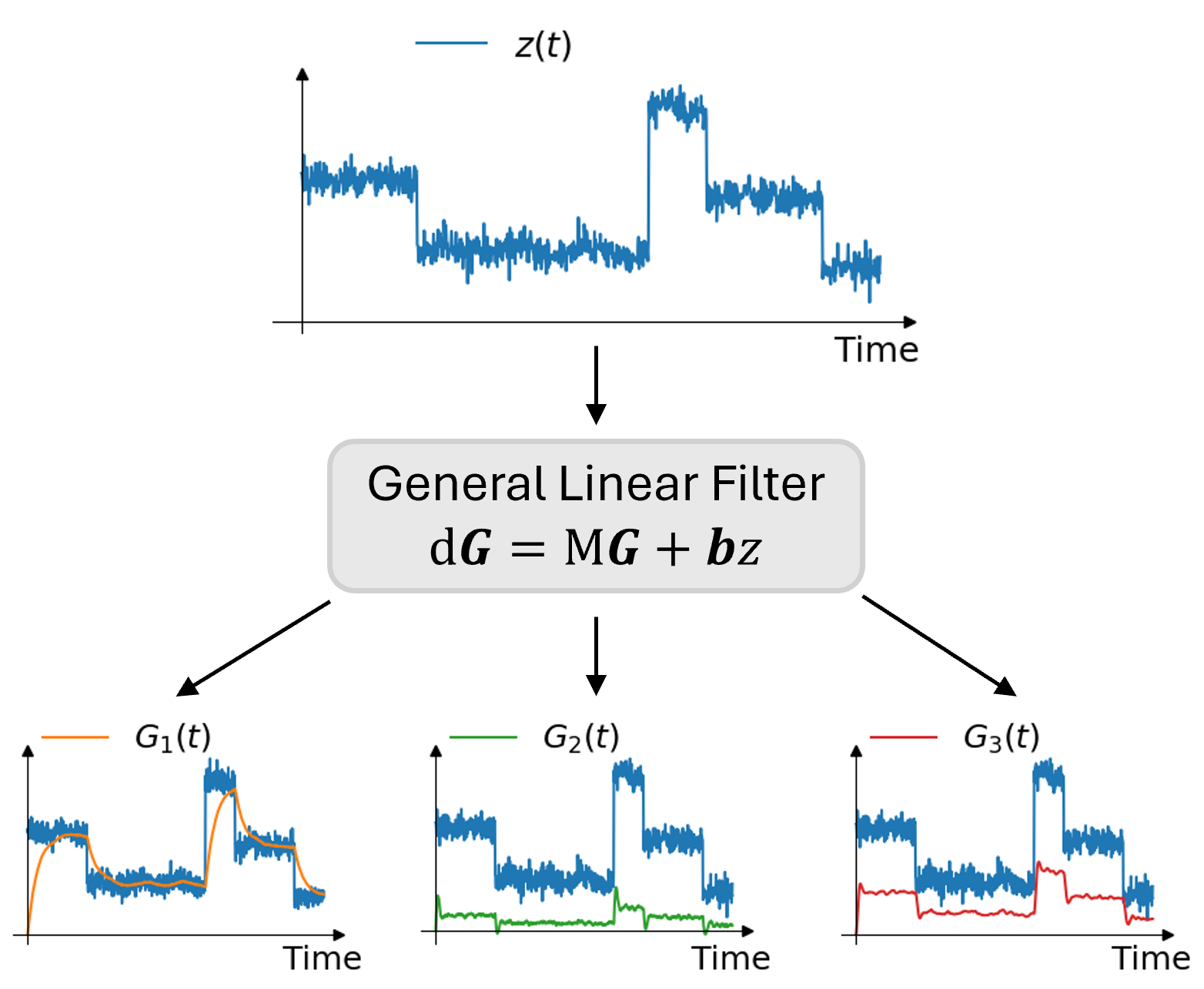}
    \caption{Illustration of the signal processing leading to the measurement outcome.
    The underlying quantum dynamics, possibly with feedback, yield a noisy measurement outcome given by $z(t)$.
    The noisy signal is then processed (filtered) by different channels, each contributing to a different component of the overall signal.
    The information processing is modeled using a general linear system.}
    \label{fig:G-signal}
\end{figure}

Once we consider the limit $\delta t \rightarrow \dd t \rightarrow 0$, we note the measurement outcome $z$ can be described by Eq.~\eqref{eq:signal_z}, with $\ddW$ being a Wiener increment that satisfies $\ddW^2 = \ddt$.
Using Ito's stochastic calculus formalism, the infinitesimal evolution of the density matrix is given by the Belavkin equation~\cite{Belavkin1989,Belavkin1989_2,Jacobs2006,Binder2019}:
\begin{equation}
    \label{eq:belavkin}
    \begin{aligned}
        d\hat{\rho}_c = \frac{-i}{\hbar}[\hat{H},\hat{\rho}_c]\dd t & + \lambda \mathcal{D}[\hat{A}] \hat{\rho}_c \dd t + \sqrt{\lambda} \dd W \{ \hat{A} - \langle \hat{A} \rangle, \hat{\rho}_c \}.
    \end{aligned}
\end{equation}
Here, $\hat{\rho}_c$ denotes the density matrix conditioned on observing the measured signal $z(t)$.
The parameter $\lambda$ characterizes the measurement strength, and controls how much backaction is caused in the system; the superoperator $\mathcal{D}[\hat{c}]\hat{\rho} = \hat{c} \hat{\rho} \hat{c}^\dagger - \frac{1}{2}\hat{c}\hat{c}^\dagger \hat{\rho} - \frac{1}{2} \hat{\rho} \hat{c} \hat{c}^\dagger$ is called dissipator.
The last term of the right-hand side of Eq.~\eqref{eq:belavkin} contains the stochastic evolution; $\langle \hat{A} \rangle = \Tr[\hat{A}\hat{\rho}_c]$ is the expected value of the measured operator, where $\{ \cdot, \cdot \}$ is the anticommutator.

At this stage, we have a continuously monitored system evolving under Eq.~\eqref{eq:belavkin} with measurement outcome given by Eq.~\eqref{eq:signal_z}.
However, the signal given by a Gaussian white noise has an ill-defined second (or higher) moment, which only allows us to apply linear feedback of the form
\begin{equation}
    \hat{H} = \hat{H}_0 + z\hat{V}.
\end{equation}
To include non-linear feedback, the authors in Ref.~\cite{AnnbyAndersson2022} introduce a new signal $D(t)$, which is the exponentially filtered version of the noisy signal $z(t)$.
The new measurement outcome $D(t)$ is continuous and better models a real physical experiment in which the information is processed at a finite rate, with a finite bandwidth.
The bandwidth is described by the rate $\gamma$ of the low-pass filter.

The signal $D(t)$ in Eq.~\eqref{eq:signal_D} can be re-written as a one-dimensional linear stochastic ODE:
\begin{align}
    \label{eq:dD}
    &\ddD{(t)} = \gamma \left( \langle \hat{A} \rangle - D{(t)} \right) \dd t + \frac{\gamma}{\sqrt{4\lambda}}\dd W.
\end{align}
The linearity of Eq.~\eqref{eq:dD} is a key ingredient to derive the QFPME in Refs.~\cite{AnnbyAndersson2022,DeSousa2025}.

Motivated by the linear model of the low-pass filter, we propose a generalization of the QFPME by considering a set $\{ G_k(t) \}_{k=1}^n$ of $n$ signals that are related by a linear system of stochastic ODEs:
\begin{equation}
    \begin{aligned}
    \label{eq:signal_G}
            &\dd{\bf G}(t) = {\rm M} {\bf G}(t) + {\bf b} z(t),
    \end{aligned}
\end{equation}
with, ${\bf G} = (G_1,\dotsc,G_n)^T$.
The matrix $M_{kj}$ and the vector ${\bf b} = (b_1,\dotsc,b_n)^T$ specify a particular model of the {processing of the measurement signal}.
As it will be clear in the next section, with examples of particular models, the signal ${\bf G}$ corresponds to a multi-dimensional processing of the noisy signal $z(t)$, as depicted in Fig.~\ref{fig:G-signal}.


The signal ${\bf G}(t)$ can be used to apply nonlinear feedback by the action of changing the Hamiltonian $\hat{H}(t) = \hat{H}({\bf G}(t))$ as a function of the measurement outcome.
The function $\hat{H}({\bf G})$ defines the feedback, i.e. a pre-determined rule on how the dynamics will be modified based on the measurement signal.

To connect the evolution of the measurement outcome(s) signal(s) to the quantum dynamics in Eq.~\eqref{eq:belavkin}, let's define the joint distribution of the quantum state and the measurement signals over the ensemble of the noise:
\begin{equation}
    \label{eq:rho_G}
    \hat{\rho}({\bf G},t) = \mathrm{E}_W[\hat{\rho}_c \delta ({\bf G}-{\bf G}{(t)})].
\end{equation}
The evolution of Eq.~\eqref{eq:rho_G} is given by:
\begin{equation}
    \label{eq:QFPME_G}
    \begin{aligned}
        &\partial_t{\hat \rho}({\bf G}) = \frac{-i}{\hbar}\left[ \hat{H}({\bf G}), \hat{\rho}({\bf G}) \right] + \lambda\mathcal{D}[\hat{A}]\hat{\rho}({\bf G}) \\
        &- \nabla_{\bf G} \cdot \left[ \frac{1}{2} \left\{ {\bf b}\hat{A} + M{\bf G}, \hat\rho({\bf G}) \right\} \right] + \frac{1}{2}\nabla_{\bf G} \cdot \left( \frac{{\bf b}{\bf b}^T}{4\lambda} \nabla_{\bf G} \hat\rho({\bf G}) \right).
    \end{aligned}
\end{equation}
See Appendix \ref{sec:appendix_general_qfpme} for a derivation of Eq.~\eqref{eq:QFPME_G}. Note the similarity of Eq.~\eqref{eq:QFPME_G} with Eq.~(18) in Ref.~\cite{Tilloy2024} {and Eq.~(1) in Ref.~\cite{Oppenheim2022}}, here derived in the context of signal processing, as will be discussed in detail in the next section.

Equation~\eqref{eq:QFPME_G} describes the evolution of the joint distribution of quantum states and measurement outcomes during the continuous measurement of the operator $\hat{A}$ and feedback $\hat{H}({\bf G})$.
The joint distribution $\hat{\rho}({\bf G},t)$ describes the ensemble of many realizations of the experiment, where, in each realization, the system is measured continuously and feedback is applied in real-time.

The first term in Eq.~\eqref{eq:QFPME_G} represents the unitary quantum dynamics generated by the total Hamiltonian $\hat{H}({\bf G})$, which can contain feedback.
The second term is the usual backaction that reduces coherences in the basis of the measured operator $\hat{A}$.
The remaining terms describe the dynamics of the noisy measurement outcomes; the third corresponds to the deterministic component of Eq.~\eqref{eq:signal_G}, which is the drift term for the Fokker-Planck equation. The fourth term represents the diffusion of the noisy signals due to the measurement, where ${\bf b}{\bf b}^T / 4\lambda$ is the general diffusion tensor for the Brownian motion in ${\bf G}$-space.

\section{Modeling various filters}
\label{sec:modeling_filters}

\begin{table*}[!ht]
    \centering
    \begin{tabular}{c@{\hspace{5mm}}c@{\hspace{5mm}}c}
    \hline\hline
        Signal & Filter model & Dimensionality\\
        \hline\rule{0pt}{2.6ex}
        ${\bf D} = (D_1,\dotsc,D_n)^T$ & Cascade of low-pass & Number of low-pass components $n$\\
        ${\bf E} = (E_1,E_2)^T$ & Band-pass & 2\\
        ${\bf F} = (F_1,\dotsc,F_n)^T$ & General kernel & Highest order of $f(t)$ ODE\\
        ${\bf G} = (G_1,\dotsc,G_n)^T$ & Arbitrary linear filter & Arbitrary\\
        \hline\hline
    \end{tabular}
    \caption{Signal notation for the different filtering models for Eq.~\eqref{eq:signal_G}.}
    \label{tab:filter_notation}
\end{table*}

In this section, we show three concrete examples of signal filtering models with increasing complexity: a sequence of low-pass filters, a band-pass filter, and a general kernel filter~{\cite{Bechhoefer2005,Oppenheim1996}}.
All models are particular examples of the linear system described by Eq.~\eqref{eq:signal_G}.
For the sake of notation, we denote the low-pass filters by the vector ${\bf D}$, the band-pass filter components by ${\bf E}$, and the general kernel filter components by ${\bf F}$, as shown in Table~\ref{tab:filter_notation}.

\subsection{Multiple low-pass filters}\label{sec:qfpme_multiple_low_pass}
Here, we will imagine that the experimental apparatus comprises $n$ filters that are applied sequentially.
This framework aims to model experiments composed of various pieces of equipment, where the output of a given layer serves as input for the next one.
For simplicity, we consider that each equipment adds an exponential low-pass filter to the signal.
The first filter is denoted by $D_1$ and is defined by Eq.~\eqref{eq:signal_D}, and the next ones are labeled $D_k$, where $k = 2,\dotsc,n$.
A low-pass filter models each additional filter
\begin{align}
    \label{eq:signal_Dk}
    &D_k(t) = \gamma_k \int_{-\infty}^t \dd s e^{-\gamma_k(t-s)} D_{k-1}(s), \quad k\geq 2 \\
    \label{eq:dDk}
    &\dd D_k{(t)} = \gamma_k \left(  D_{k-1}{(t)} - D_k{(t)} \right) \dd t.
\end{align}
The parameters $\{\gamma_k\}$, represent the experimental bandwidth of the filters.
After all layers of filtering are taken into account, we can apply time-dependent feedback by changing the Hamiltonian in terms of these delayed signals $\hat{H} = \hat{H}({\bf D}(t))$, where ${\bf D}{(t)} = (D_1{(t)},\dotsc,D_n{(t)})$ represents the set of all filters.

Using the framework described in Sec.~\ref{sec:qfpme_general}, we can write Eqs.~(\ref{eq:dD},\ref{eq:dDk}) as a linear system of the form
\begin{equation}
    \label{eq:matrix_M_low-pass}
    {\rm M} = \begin{pmatrix}
        -\gamma_1 & 0 & 0& \dotsc & 0\\
        \gamma_2 & -\gamma_2 & 0 & \dotsc & 0\\
        0 & \gamma_3 & -\gamma_3 & \dotsc & 0 \\
        \vdots & \vdots & & \ddots & \vdots \\
        0 & & \dotsc & \gamma_n & -\gamma_n
    \end{pmatrix} \quad\quad {\bf b} = \begin{pmatrix}
        \gamma_1 \\ 0 \\ \vdots \\ 0 \\ 0
    \end{pmatrix},
\end{equation}
which yields a corresponding QFPME for multiple low-pass filters
\begin{equation}
    \begin{aligned}
    \label{eq:qfpme_filters}
        &\partial_t\hat{\rho}({\bf D}) = 
        \frac{-i}{\hbar}\left[ \hat{H}({\bf D}), \hat{\rho}({\bf D}) \right] \\
        &+ \lambda\mathcal{D}[\hat{A}]\hat{\rho}({\bf D}) -\frac{\gamma_1}{2}\partial_{D_1}\{ \hat{A}-D_1, \hat{\rho}({\bf D}) \}\\
        &-\sum_{k=2}^{n} \gamma_k \partial_{D_k} [(D_{k-1}-D_k)\hat{\rho}({\bf D})]  
        + \frac{\gamma^2}{8\lambda}\partial_{D_1}^2 \hat{\rho}({\bf D}).
    \end{aligned}
\end{equation}
The generalized drift represents the relaxation of the $k'$th filter with its characteristic bandwidth $\gamma_k$.
The only diffusive term in Eq.~\eqref{eq:qfpme_filters} is the signal $D_1$.
This happens because all the other signals evolve deterministically toward one another.
In contrast, the first signal evolves under an OU process towards the measured operator expectation value.

\subsection{Band-pass filtering}\label{sec:qfpme_band-pass}
We can construct the QFPME with a band-pass filter by changing the form of the filtering operation in Eq.~\eqref{eq:signal_D}.
There are multiple ways to implement a band-pass filter, i.e. a filter that only allows signals in an intermediate range of frequencies in the spectral domain.
Here, we will focus on a particular realization of band-pass that takes the form of a Lorentzian function in the Fourier domain.
This filter can be written as a frequency shift of the exponential low-pass filter~{\cite{Oppenheim1996}} in Eq.~\eqref{eq:signal_D},
\begin{equation}
    \label{eq:signal_E1}
    E_1(t) = \gamma \int_{-\infty}^t \dd s e^{-\gamma(t-s)} \cos{(\Omega(t-s))} z(s).
\end{equation}
The signal in Eq.~\eqref{eq:signal_E1} obeys a second-order equation, so we need to define the quadrature $E_2(t)$:
\begin{equation}
    \label{eq:signal_E2}
    E_2(t) = \gamma \int_{-\infty}^t \dd s e^{-\gamma(t-s)} \sin{(\Omega(t-s))} z(s).
\end{equation}
The following system of equations couples the two signals that constitute the band-pass filter:
\begin{equation}
    \begin{aligned}
        \label{eq:dE}
        &\dd E_1{(t)} = \gamma \left( z - E_1{(t)} \right)\dd t - \Omega E_2{(t)} \dd t, \\
        &\dd E_2{(t)} = \left(\Omega E_1{(t)} - \gamma E_2{(t)}\right) \dd t,
    \end{aligned}
\end{equation}
which in matrix form can be written as
\begin{equation}
    \label{eq:matrix_M_band-pass}
    {\rm M} = \begin{pmatrix}
        -\gamma & -\Omega \\
        \Omega & -\gamma
    \end{pmatrix} \quad\quad {\bf b} = \begin{pmatrix}
        \gamma \\ 0
    \end{pmatrix}.
\end{equation}

Combining Eqs.~(\ref{eq:QFPME_G},\ref{eq:matrix_M_band-pass}), we get the QFPME for a band-pass filter
\begin{equation}
    \begin{aligned}
    \label{eq:qfpme_band-pass}
        &\partial_t\hat{\rho}({\bf E}) = 
        \frac{-i}{\hbar}\left[ \hat{H}({\bf E}), \hat{\rho}({\bf E}) \right] + \lambda\mathcal{D}[\hat{A}]\hat{\rho}({\bf E}) \\
        &-\partial_{E_1} \left[ \frac{1}{2}\left\{ \gamma\hat{A} - \gamma E_1 -\Omega E_2, \hat{\rho}({\bf E}) \right\} \right]  \\
        &- \partial_{E_2}\left[(\Omega E_1 - \gamma E_2)\hat{\rho}({\bf E})\right] + \frac{\gamma^2}{8\lambda}\partial_{E_1}^2 \hat{\rho}({\bf E}),
    \end{aligned}
\end{equation}
where ${\bf E} = (E_1,E_2)$.
The evolution of the signals $(E_1(t), E_2(t))$ are captured by the drift term in Eq.~\eqref{eq:qfpme_band-pass}, and the diffusion is only present in $E_1$ because it is the one that has a stochastic term from the measurement outcome $z(t)$.

In contrast to the low-pass filter, the band-pass filter is described by the two quadratures $(E_1(t), E_2(t))$ that interact under the dynamics given by Eq.~\eqref{eq:dE}.
Thus, to fully describe the system, one needs to solve the QFPME in a two-dimensional space.
If the quadrature $E_2(t)$ is not experimentally accessible, one needs to integrate out its hidden degrees of freedom to get an effective equation. 
This effective equation is similar to having a memory kernel, or a non-Markovian description of the system.


\subsection{General kernel filtering}
Motivated by the construction in Sec.~\ref{sec:qfpme_band-pass}, we can write how the QFPME changes when we use a general filter of the form
\begin{equation}
    \label{eq:general_filter}
    F_1(t) = \int_{-\infty}^t \dds f(t - s) z(s).
\end{equation}
The function $f(t)$ is the filter kernel. We consider the case where $f(t)$ obeys a finite ODE~{\cite{Oppenheim1996}}:
\begin{equation}
    \label{eq:gen_filter_ode}
    f^{(n)}(t) + a_{n-1}f^{(n-1)}(t) + \cdots + a_0 f(t) = 0.
\end{equation}
The low-pass filter considered in Sec.~\ref{sec:qfpme_band-pass} obeys a first-order ODE, and the band-pass filter considered in Sec.~\ref{sec:qfpme_band-pass} obeys a second-order ODE.

To solve for the  QFPME associated with the filter in Eq.~\eqref{eq:general_filter}, we need to define $n-1$ components of the filter
\begin{equation}
\label{eq:gen_aux_signal}
    F_{k}{(t)} = \int_{-\infty}^t \dd s f^{(k-1)}(t - s) z(s), \quad 2 \leq k \leq n.
\end{equation}
These signals evolve under the following coupled ODE:
\begin{equation}
\label{eq:ODE_general_analytic_filter}
    \begin{aligned}
        &\dd F_1{(t)} = f(0)z(t)\dd t + F_2{(t)}\dd t \\
        &\dd F_{k+1}{(t)} = f^{(k)}(0)z(t)\dd t + F_{k+2}{(t)}\dd t\\
        &\dd F_{n}{(t)} = f^{(n-1)}(0)z(t)\dd t - \sum_{j=1}^{n}a_{j-1} F_{j}{(t)}\dd t.
    \end{aligned}
\end{equation}
In matrix form,
\begin{equation}
    \label{eq:matrix_M_general_filter}
    {\rm M} = \begin{pmatrix}
        0 & 1 & 0& \dotsc & 0\\
        0 & 0 & 1 & \dotsc & 0\\
        \vdots & \vdots & & \ddots & \vdots \\
        -a_0 & -a_1 & -a_2 & \dotsc & -a_{n-1}
    \end{pmatrix} \quad\quad {\bf b} = \begin{pmatrix}
        f(0) \\ f^{(1)}(0) \\ \vdots  \\ f^{(n-1)}
    \end{pmatrix}.
\end{equation}

Combining Eqs.~(\ref{eq:QFPME_G},\ref{eq:matrix_M_general_filter}) we have the following \mbox{QFPME}:
\begin{equation}
    \label{eq:QFPME_analytic_general_filter}
    \begin{aligned}
        &\partial_t \hat{\rho}({\bf F}) = \frac{-i}{\hbar}\left[ \hat{H}({\bf F}), \hat{\rho}({\bf F}) \right] + \lambda \mathcal{D}[\hat{A}]\hat{\rho}({\bf F})\\
        &-\sum_{j=1}^{n-1} \partial_{F_{j}} \left[ \frac{1}{2}\left\{f^{(j-1)}(0) \hat{A} + F_{j+1}, \hat{\rho}({\bf F}) \right\} \right] \\
        &-\partial_{F_n} \left[ \frac{1}{2}\left\{ f^{n-1}(0)\hat{A} - \sum_{k=1}^n a_{k-1}F_k , \hat{\rho}({\bf F}) \right\} \right]\\
        &+\sum_{j=1}^n\sum_{k=1}^n \frac{f^{(j-1)}(0) f^{(k-1)}(0)}{8\lambda} \partial_{F_j}\partial_{F_k} \hat{\rho}({\bf F}),
    \end{aligned}
\end{equation}
where ${\bf F} = (F_1, F_2, \dotsc, F_n)$. The drift terms describe the deterministic evolution of each signal, where the final signal is given by the ODE that defines the filter.
The second derivative can contain cross terms depending on how many signals have stochastic evolution in Eq.~\eqref{eq:ODE_general_analytic_filter}.

This general QFPME for the $f(t)$ kernel shows that as we increase the complexity of the signal processing, one needs to enlarge the state space of the filter.
The dimension of the ${\bf F}$ is the same as the order of the ODE, showing that the evolution of the system remains Markovian in the large space when we take all components of the filter into account.
However, if one of the components $F_k(t)$ is not accessible for the experiment, one can integrate it out to get an effective description of the system.
The resulting dynamics will contain non-Markovian effects that stem from the reduction of dimensionality.

\section{Cooling a harmonic oscillator}
\label{sec:harmonic-oscillator}

This section focuses on applying the general QFPME  to the toy model used in Ref.~\cite{DeSousa2025}.
See Appendix \ref{sec:appendix_cooling} for details on deriving the results presented in this section.

Consider a quantum harmonic oscillator of mass $m$ and characteristic frequency $\omega$ undergoing continuous measurement and feedback. Introducing dimensionless operators $\hat{p} \rightarrow \hat{p}\sqrt{\hbar m \omega}$ and $\hat{x} \rightarrow \hat{x}\sqrt{\hbar/m\omega}$, the Hamiltonian in the absence of feedback becomes 
\begin{equation}
    \label{eq:bare_hamiltonian}
    \hat{H}_0 = \frac{\hbar\omega}{2}\left( \hat{p}^2 + \hat{x}^2 \right).
\end{equation}
To apply feedback, we consider that position and momentum are being continuously monitored, and the Hamiltonian is changed based on both processed measurement signals.
The general QFPME for both $\hat{x}$ and $\hat{p}$ measurement is given by
\begin{equation}
    \label{eq:QFPME_QHO}
    \begin{aligned}
        &\partial_t \hat{\rho}({\bf G}_x,{\bf G}_p) = \frac{-i}{\hbar}\left[ \hat{H}({\bf G}_x,{\bf G}_p), \hat{\rho}({\bf G}_x,{\bf G}_p) \right] \\
        &+ \lambda \mathcal{D}[\hat{x}]\hat{\rho}({\bf G}_x,{\bf G}_p) + \lambda \mathcal{D}[\hat{p}]\hat{\rho}({\bf G}_x,{\bf G}_p)\\
        &+\FF[{\rm M},{\bf b},\hat{x}]\hat\rho({\bf G}_x,{\bf G}_p) + \FF[{\rm M},{\bf b},\hat{p}]\hat\rho({\bf G}_x,{\bf G}_p),\\
    \end{aligned}
\end{equation}
with,
\begin{equation}
    \label{eq:FF_QHO}
    \begin{aligned}
        &\FF[{\rm M},{\bf b},\hat{\xi}]\hat\rho({\bf G}_x,{\bf G}_p) = \\
        &-\nabla_{{\bf G}_{\xi}} \left[ \frac{1}{2}\left\{{\bf b}\hat{\xi} + M {\bf G}_{\xi}, \hat{\rho}({\bf G}_x,{\bf G}_p) \right\} \right] \\
        &+ \frac{1}{2} \nabla_{{\bf G}_{\xi}} \cdot \left( \frac{{\bf b}{\bf b}^T}{4\lambda}\nabla_{{\bf G}_{\xi}} \hat{\rho}({\bf G}_x,{\bf G}_p) \right).
    \end{aligned}
\end{equation}

The task we are interested in this particular example is to cool the particle to its lowest possible energy.
The feedback we focus on is similar to the cooling feedback used in Ref.~\cite{DeSousa2025}, where the Hamiltonian changes by moving the center of the position and momentum harmonic traps to the measurement outcome.
The goal is to keep the particle at the bottom of the potential and reduce its energy.

{Our focus on this application is to study the evolution of the ensemble average steady state energy of the harmonic oscillator under feedback with different signal processing.
Formally, we define our quantity of interest as}
\begin{equation}
    \begin{aligned}
        \label{eq:definition_average_energy}
        {\sev{\hat{H}}}~&{= \int \dd{\bf G_x}\dd{\bf G_p} \tr[\hat{H}({\bf G_x}, {\bf G_p}) \hat{\rho}({\bf G_x},{\bf G_p},t)]} \\
        &{= \mathrm{E}_W[\hat{H}({\bf G_x}, {\bf G_p}) \hat\rho_c]}
    \end{aligned}
\end{equation}
{To solve for the energy steady state, we calculate the energy time derivative $\partial_t \sev{\hat H}$ (and other relevant quantities) by plugging Eq.~\eqref{eq:QFPME_QHO} into Eq.~\eqref{eq:definition_average_energy}.
This procedure holds for all protocols studied in this section.}

\subsection{One layer of filtering}\label{sec:QHO_1layer_filtering}
To apply cooling using a single layer of low-pass filter, we use the original QFPME model~\cite{AnnbyAndersson2022,DeSousa2025}, which is a particular case of Eq.~\eqref{eq:matrix_M_low-pass} with $n= 1$.
The feedback Hamiltonian is given by:
\begin{equation}
    \label{eq:H1}
    \hat{H}_1({\bf D_1}) = \frac{\hbar\omega}{2}\left[ \left( \hat{p}-D_{p,1} \right)^2 + \left( \hat{x}-D_{x,1} \right)^2 \right],
\end{equation}
where ${\bf D_1} = (D_{x,1},D_{p.1})$.

As shown in Ref.~\cite{DeSousa2025}, the average energy evolves by the following equation
\begin{equation}
    \label{eq:ode_H1}
    \partial_t \sev{\hat{H}_1} + 2\gamma \sev{\hat{H}_1} = 2\gamma \sev{\hat{H}_1}_\infty,
\end{equation}
where
\begin{equation}
    \label{eq:H1-infty}
    \sev{\hat{H}_1}_\infty \equiv \lim_{t\rightarrow\infty} \sev{\hat{H}_1}(t) = \frac{\hbar\omega}{2}\left( \frac{\lambda}{\gamma} + \frac{\gamma}{4\lambda} \right).
\end{equation}

This simple protocol can cool down the particle to ground state if we choose the parameters $\gamma = 2\lambda$. In this case, we have 
\begin{equation}
    \sev{\hat{H}_1}_\infty = \frac{\hbar \omega}{2}.
\end{equation}

\subsection{Two layers of filtering}
Here, we consider two layers of filtering, i.e. $n = 2$ in Eq.~\eqref{eq:matrix_M_low-pass}.
We will focus on the feedback Hamiltonian given by
\begin{equation}
    \label{eq:H2}
    \hat{H}_2({\bf D_2}) = \frac{\hbar\omega}{2}\left[ \left( \hat{p}-D_{p,2} \right)^2 + \left( \hat{x}-D_{x,2} \right)^2 \right].
\end{equation}
This means the Hamiltonian is changed instantaneously based on a double filtered signal $D_{x/p,2}$. The filtering frequencies are $\gamma_1 = \gamma$ and $\gamma_2 = \Omega$, see Eqs.~\eqref{eq:signal_D} and \eqref{eq:signal_Dk}. For the sake of notation, let's denote $U_2 \equiv \sev{\hat{H}_2(\bf {D_2})}$.

Using Eqs.~\eqref{eq:qfpme_filters} and \eqref{eq:H2}, we can write a single ordinary differential equation for the average energy of the double-filtered protocol:
\begin{equation}
    \begin{aligned}
    \label{eq:ode_H2}
        &\partial_t^4 U_2 + 4(\Omega+\gamma)\partial_t^3 U_2 + [4\Omega\gamma + 5(\Omega+\gamma)^2 + \omega^2]\partial_t^2 U_2 \\
        &+ 2(\Omega+\gamma)[4\Omega\gamma + (\Omega+\gamma)^2+\omega^2]\partial_t U_2 \\
        &+ 4\Omega\gamma (\Omega+\gamma)^2 [U_2 - U_{2,\infty}] = 0,
    \end{aligned}
\end{equation}
where 
\begin{equation}
    \label{eq:H2-infty}
    U_{2,\infty} = \frac{\hbar\omega}{2}\left[ \frac{\lambda}{\tilde{\gamma}} + \frac{\tilde{\gamma}}{4\lambda} + \frac{\lambda}{(\Omega+\gamma)} +\frac{\lambda\omega^2}{\tilde{\gamma}(\Omega+\gamma)^2} \right],
\end{equation}
and
\begin{equation}
    \label{eq:effective_frequency}
    \frac{1}{\tilde{\gamma}} = \frac{1}{\gamma} + \frac{1}{\Omega}
\end{equation}
is the effective frequency of the filter. The asymptotic solution for the average energy that evolves under Eq.~\eqref{eq:ode_H2} is given by
\begin{equation}
    \sev{\hat{H}_2}_\infty = \lim_{t \rightarrow \infty} U_2(t) = U_{2,\infty}.
\end{equation}

One can recover the results of the single filtered protocol in the limit $\Omega \gg \gamma$. In this regime, the evolution of the average energy in Eq.~\eqref{eq:ode_H2} can be written in the following form:
\begin{equation}
    \begin{aligned}
    \label{eq:ode_H2_approx}
        &\frac{5}{2\Omega}\partial_t^2 U_2 + \left( 1 + \frac{5\gamma}{\Omega} \right)\partial_t U_2 + \left( 2\gamma + \frac{4\gamma^2}{\Omega} \right) U_2 \\
        &= \hbar \omega \left( \lambda + \frac{\gamma^2}{4\lambda} \right) + \frac{\hbar \omega}{\Omega}\left( 4\gamma\lambda + \frac{\gamma^3}{4\lambda} \right).
    \end{aligned}
\end{equation}
In this limit, the system will relax to the following asymptotic energy
\begin{equation}
    \begin{aligned}
        \label{eq:H2-infty_approx}
        \sev{\hat{H}_2}_\infty = \frac{\hbar \omega}{2}\left( \frac{\lambda}{\gamma} + \frac{\gamma}{4\lambda} \right) + \frac{\hbar \omega}{\Omega}\left( \lambda - \frac{\gamma^2}{8\lambda} \right).
    \end{aligned}
\end{equation}
Note that the Eqs.~\eqref{eq:ode_H2_approx} and \eqref{eq:H2-infty_approx} are equal to the Eqs.~\eqref{eq:ode_H1} and \eqref{eq:H1-infty}, respectively, up to zeroth order in $1/\Omega$.

While Eq.~\eqref{eq:H2-infty} reveals that no choice of parameters achieves ground state cooling, we can use Eq.~\eqref{eq:H2-infty_approx} to cool down the particle to a lower energy than the single layer of filter protocol if our small correction is negative. The first order correction of Eq.~\eqref{eq:H2-infty_approx} is negative if
\begin{equation}
    \label{eq:H2_correction_condition}
    \frac{\gamma}{\lambda} > 2\sqrt{2}.
\end{equation}

These results show an interesting feature of this system.
We can achieve better cooling by introducing another filtering layer on top of the original noisy signal if the parameters $\gamma$ and $\lambda$ satisfy Eq.~\eqref{eq:H2_correction_condition}.

\subsection{Three layers of filtering}
For three layers of filtering, we have $n = 3$ in Eq.~\eqref{eq:matrix_M_low-pass}.
The feedback Hamiltonian given by
\begin{equation}
    \label{eq:H3}
    \hat{H}_3({\bf D_3}) = \frac{\hbar\omega}{2}\left[ \left( \hat{p}-D_{p,3} \right)^2 + \left( \hat{x}-D_{x,3} \right)^2 \right].
\end{equation}
For the bandwidth frequencies we choose $\gamma_1 = \gamma$, $\gamma_2 = \Omega$, and $\gamma_3 = \Omega$.
Now, the system of equations necessary to solve for the energy is substantially more complex (see Appendix \ref{sec:appendix_cooling}).
We have nine linearly coupled equations to solve, so we will only focus on the asymptotic energy.

The asymptotic energy for cooling protocol with three layers of filtering is given by
\begin{equation}
    \label{eq:H3-infty}
    \begin{aligned}
    &\sev{\hat{H}_3}_\infty = \\
        &\frac{\frac{\hbar \omega}{2}
        \left(
        \begin{aligned}
            8\lambda^2\Omega^3(\omega^2+\Omega^2)^2 \\
            + 8\gamma\lambda^2\Omega^2(3\omega^4+7\omega^2\Omega^2+8\Omega^4) \\
            + \gamma^5[\Omega^4+4\lambda^2(\omega^2+5\Omega^2)] \\
            + 2\gamma^4[2\Omega^5+\lambda^2(9\omega^2\Omega+48\Omega^3)] \\
            + \gamma^3[5\Omega^6+4\lambda^2(\omega^4+10\omega^2\Omega^2+45\Omega^4)] \\
            + 2\gamma^2[\Omega^7+\lambda^2(9\omega^4\Omega+31\omega^2\Omega^3+80\Omega^5)]
        \end{aligned}
        \right)}{
        2\gamma\lambda\Omega^2 \left(
        \begin{aligned}
            4\gamma^4\Omega-4\omega^2\Omega^3+4\Omega^5 - 2\gamma^3(\omega^2-8\Omega^2) \\
            + \gamma^2(-9\omega^2\Omega+24\Omega^3) \\
            + 4\gamma(-3\omega^2\Omega^2+4\Omega^4)
        \end{aligned}
        \right)}.
    \end{aligned}
\end{equation}
Because the system is hard to solve analytically, we don't have the exact solution for which choices of parameters $(\omega,\gamma,\lambda,\Omega)$ allow for the system to relax to the solution in Eq.~\eqref{eq:H3-infty}.
Some choices of parameters yield unphysical results when applied to Eq.~\eqref{eq:H3-infty}, which correspond to parameter regime where the system is being heated indefinitely.

If we focus on the limit $\Omega \gg \omega,\gamma,\lambda$, we expect the solution to be close to the single-layer protocol. In the limit $\Omega \rightarrow \infty$ we get
\begin{equation}
    \begin{aligned}
        \label{eq:H3-infty_approx}
        \sev{\hat{H}_3}_\infty = \frac{\hbar \omega}{2}\left( \frac{\lambda}{\gamma} + \frac{\gamma}{4\lambda} \right) + \frac{\hbar \omega}{\Omega}\left( 2\lambda - \frac{3\gamma^2}{16\lambda} \right).
    \end{aligned}
\end{equation}
Note that this protocol yields lower asymptotic energy than the single-layer protocol in Eq.~\eqref{eq:H1-infty}, if
\begin{equation}
    \frac{\gamma}{\lambda} > 2\sqrt{\frac{8}{3}}\lambda.
\end{equation}

\begin{figure}[t!]
\centering
  \includegraphics[width=1.0\linewidth]{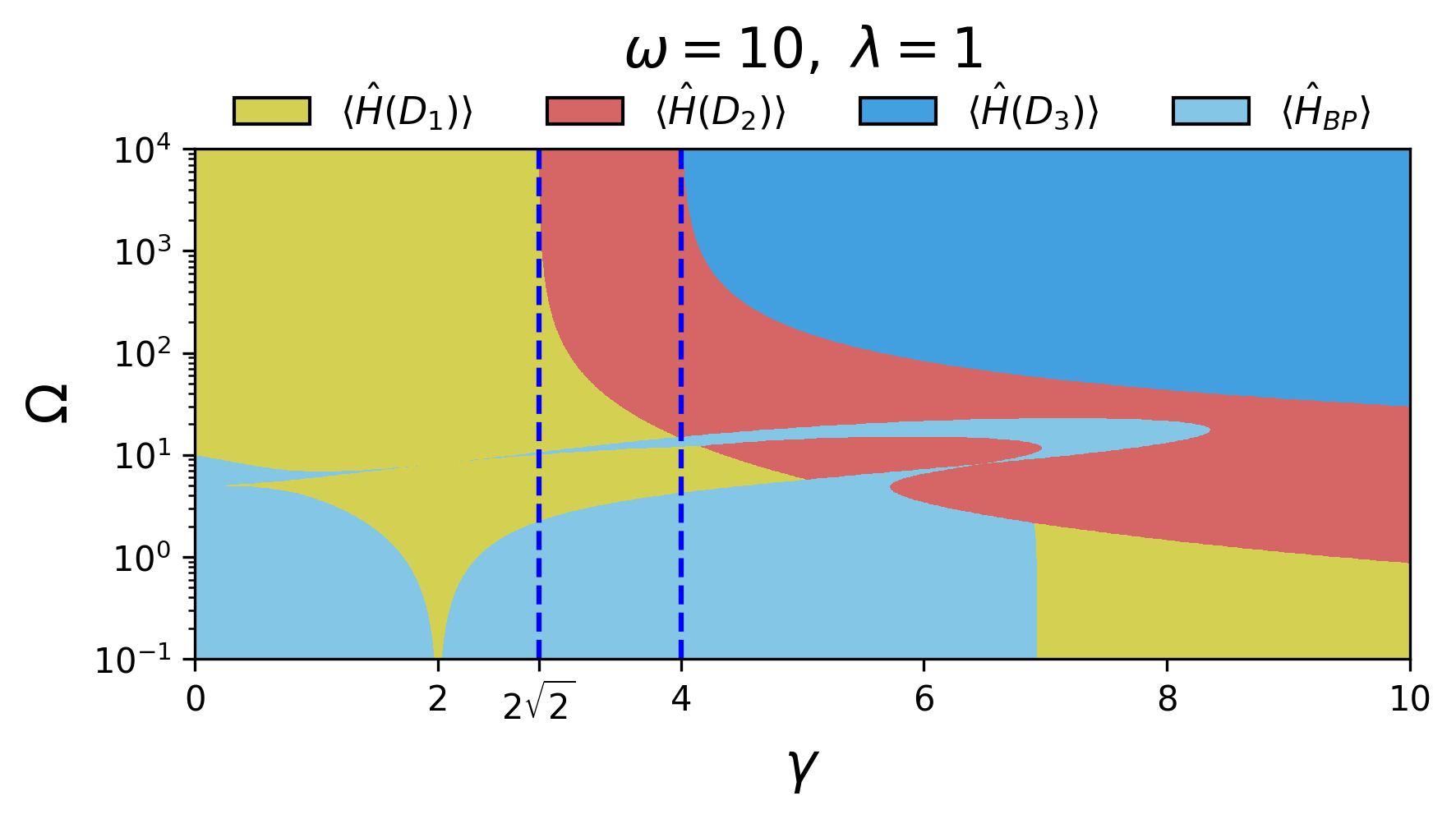}
  \caption{Phase diagram indicating which protocol is able to cool down the particle to the lowest energy for a given value of $(\gamma,\Omega)$. Each region indicates which protocol yields the lowest energy using Eqs.~(\ref{eq:H1-infty},\ref{eq:H2-infty},\ref{eq:H3-infty},\ref{eq:band_pass_energy}). The vertical dashed lines indicate the transition regions where different low-pass filters have the best cooling in the limit $\Omega \rightarrow \infty$ (see Eq.~\eqref{eq:g_l_decision_region}).}
\label{fig:qfpme_multi_filter}
\end{figure}

When we combine Eqs.~\eqref{eq:H2-infty_approx} and \eqref{eq:H3-infty_approx}, we can solve for which regions of the parameter space $(\gamma,\lambda)$ we get better cooling with each of the three protocols:
\begin{equation}
    \label{eq:g_l_decision_region}
    \Omega \rightarrow \infty: \left\{
    \begin{alignedat}{3}
        &\gamma/\lambda \leq 2\sqrt{2}: &&\min(\sev{\hat{H}_{1,2,3}}_\infty) = \sev{\hat{H}_1}_\infty\\
        &2\sqrt{2} < \gamma/\lambda \leq 4: \quad&&\min(\sev{\hat{H}_{1,2,3}}_\infty) = \sev{\hat{H}_2}_\infty\\
        &\gamma / \lambda > 4: &&\min(\sev{\hat{H}_{1,2,3}}_\infty) = \sev{\hat{H}_3}_\infty
    \end{alignedat}
    \right.
\end{equation}
Using the table in Eq.~\eqref{eq:g_l_decision_region}, one can design the experiment to artificially filter the measurement signal and apply feedback based on a $n-$filtered signal. This cleaner signal can yield better cooling if the parameters $\gamma/\lambda$ cannot be controlled easily.

\subsection{Band-pass filter}
A similar analysis can be done using the band-pass filter master equation in Eqs.~(\ref{eq:QFPME_QHO},\ref{eq:FF_QHO},\ref{eq:matrix_M_band-pass}).
The protocol is given by
\begin{equation}
    \label{eq:HE1_band_pass}
    \hat{H}_{\rm BP}({\bf E_1}) = \frac{\hbar\omega}{2}\left[ (\hat{p}-E_{p,1})^2 + (\hat{x}-E_{x,1})^2 \right].
\end{equation}
The lack of symmetries in the equations of motion for this system (see Appendix \ref{sec:appendix_cooling}) makes its solution hard to interpret. The asymptotic ensemble energy for the cooling of a quantum harmonic oscillator with band-pass filtering is
\begin{equation}
    \begin{aligned}
        \label{eq:band_pass_energy}
        \sev{\hat{H}_{\rm BP}}_\infty = &\frac{\hbar\omega}{2}\left( \frac{\lambda}{\gamma} + \frac{\gamma}{4\lambda} \right)\left[ 1+\frac{\Omega^2}{\omega^2}\frac{(4\gamma^2-\omega^2+4\Omega^2)}{(4\gamma^2+\omega^2-4\Omega^2)} \right] \\
        &+ \frac{\hbar\omega}{2}\frac{\gamma\Omega^2(3\omega^2-4\gamma^2-4\Omega^2)}{4\lambda\omega^2(4\gamma^2+\omega^2-4\Omega^2)}.
    \end{aligned}
\end{equation}

The conditions on the parameters $(\omega,\gamma,\lambda,\Omega)$ for the system of equations to converge {cannot easily be solved}.
However, it is not hard to check that for $\Omega^2 \approx \gamma^2+\omega^2/4$, Eq.~{\eqref{eq:band_pass_energy}} can yield unphysical values, namely $\sev{\hat{H}_\text{BP}}_\infty \leq \hbar\omega/2$.
The breakdown of the asymptotic solution for some parameters indicates that the protocol fails to cool down the particle to a finite value.
{Mathematically, the unphysical values in Eq.~\eqref{eq:band_pass_energy} stem from the failure of the system in Eq.~\eqref{eq:appendix_system_HBP} to converge to a steady state.}

Combining all results from Eqs.~(\ref{eq:H1-infty},\ref{eq:H2-infty},\ref{eq:H3-infty},\ref{eq:band_pass_energy}), we plot in Fig.~\ref{fig:qfpme_multi_filter} which protocol (denoted by the different colors) yields the best cooling as a function of the filter parameters.
{To plot Fig.~\ref{fig:qfpme_multi_filter}, we explicitly assume that all conditions that satisfy $\sev{H} \geq \hbar\omega/2$ are achievable, which is a simplification as we don't have the analytical solution to when some protocols converge.
A careful analysis on the convergence of these strategies is left for future work.}
The dashed lines show the results of the low-pass filters in the limit $\Omega\rightarrow\infty$.
The intricate boundaries of the band-pass filter arise from the non-trivial energy dependency on the oscillatory parameter $\Omega$.

\section{Discussion and outlook}
\label{sec:discussion}

In this work, we extended the results derived by Annby-Andersson in Ref.~\cite{AnnbyAndersson2022} by considering a generic linear filtering of the measurement outcome signal.
We derived a general master equation that can model a variety of signal filters: cascaded low-pass, resonant band-pass, or arbitrary kernels that obey certain analytical properties.
The particular cases of low-pass and band-pass filters are applied to the study of cooling of a quantum harmonic oscillator~\cite{DeSousa2025}.
Our results add to the interpretability of the master equations derived in {Refs.~\cite{Tilloy2024,Oppenheim2022}} in terms of filtering operations for experimentalists.

The general form of the QFPME can be used to model complex experiments by capturing their signal processing structure.
The master equation in Eq.~\eqref{eq:QFPME_G} exists in an enlarged filter space, where the dynamics remain strictly Markovian.
However, when only a subset of signals is experimentally accessible, integrating out hidden variables generates effective memory kernels.
This provides a natural way to derive non-Markovian feedback equations directly from our formalism.

Our master equations were also applied to a particular toy model for cooling a quantum harmonic oscillator.
We derived analytical results for the asymptotic ensemble energy for various protocols.
For quadratic Hamiltonians, it is possible to close the equations of motion and get a simple set of equations whose solution describes the quantum system's energy, position, and momentum.
The analytical results show that changing the filter can have a non-trivial influence on the system's cooling capacity.

The optimal cooling happens for a single low-pass filter at $\gamma/\lambda = 2$.
Nevertheless, away from this condition, it can be beneficial to change the spectral properties of the filtering to improve cooling, as seen in our phase diagram in Fig.~\ref{fig:qfpme_multi_filter}.
By considering changes in the signal processing, we have a larger parameter space in which the experimentalists can apply optimization techniques to enhance the performance of quantum feedback protocols.

We anticipate our cooling model being used to study optomechanical systems.
The protocols and the signal filterings can be readily applied to standard platforms, such as optical tweezers.
Beyond cooling, our framework could aid in optimizing feedback for entanglement stabilization or squeezing protocols.
We hope these results motivate both theorists and experimentalists to adopt ensemble-level master equations as tools for designing robust feedback systems.

\vspace{1mm}
\noindent
{\bf Acknowledgments.}
The author thanks Bj\"orn Annby-Andersson and Christopher Jarzynski for inspiring discussions.
This work was funded by the Quantum Leap Challenge Institute for Robust Quantum Simulation (OMA-2120757) and São Paulo Research Foundation (2024/20892-8).

\bibliography{references}

\appendix
\makeatletter
\def\@seccntformat#1{%
  \appendixname~\csname the#1\endcsname.\quad
}
\def\@sec@title@format#1#2{#1\quad #2\par}
\makeatother

\clearpage
\widetext



\section{Deriving the general 
QFPME}\label{sec:appendix_general_qfpme}
Here, we derive the most general form of the QFPME in Eq.~\eqref{eq:QFPME_G}.
The stochastic equations that one needs to consider are the Belavkin equation and the general SDE for the filters:
\begin{align}
    &\dd{\hat{\rho}}_c = \frac{-i}{\hbar}\left[ \hat{H}({\bf G}{(t)}), \hat{\rho}_c \right]\dd{t} + \lambda\mathcal{D}[\hat{A}]\hat{\rho}_c\dd{t} + \sqrt{\lambda}\dd{W}\{ \hat{A}-\sev{\hat A}, \hat\rho_c \}\\
    &\dd G_k{(t)} = b_k \left( \sev{\hat A} + \frac{1}{\sqrt{4\lambda}}\frac{\dd{W}}{\dd{t}} \right)\dd t + \sum_{j = 1}^n M_{kj} G_j{(t)}\dd t, \quad k = 1,\dotsc,n.
\end{align}
Here, the matrix $M$ and the vector ${\bf b}$ fully specify the model of the filter.

The object one needs to focus on to derive the general QFPME is the joint distribution over the trajectories (realizations of the stochastic variable $\dd{W}$) that are consistent with the measurement outcome ${\bf G}(t) = (G_1(t),\dotsc,G_n(t))$:
\begin{equation}
    \label{eq:appendix_def_rho_G}
    \hat{\rho}({\bf G},t) = \mathrm{E}_{W}\left[ \hat{\rho}_c(t)\prod_{k=1}^n \delta(G_k^{}(t) - G_k) \right] = {\rm E}_W \left[ \hat{\rho}_c {\bf \delta}({\bf G}(t) - {\bf G})\right] \equiv \mathrm{E}_W [\hat\rho_c({\bf G})].
\end{equation}
The infinitesimal evolution of Eq.~\eqref{eq:appendix_def_rho_G} can be calculated using Ito's rule of stochastic calculus:
\begin{align}
    &\begin{aligned}
        \dd{\hat \rho}({\bf G}) = {\rm E}_W
        &\left[ \dd{\hat \rho_c}\prod_{k=1}^n \delta(G_k^{}(t) - G_k) + \hat{\rho}_c\sum_k \dd\delta(G_k^{}(t) - G_k) \prod_{j\neq k}^n \delta(G_j^{}(t)-G_j) + \right. \\
        &~~\dd{\hat \rho_c}\sum_k \dd\delta(G_k^{}(t) - G_k) \prod_{j\neq k}^n \delta(G_j^{}(t)-G_j) + \\
        &~~\left.\hat{\rho}_c\sum_{k}\sum_{j\neq k} \dd\delta(G_k^{}(t) - G_k) \dd\delta(G_j^{}(t) - G_j) \prod_{i\neq k,j}^n \delta(G_i^{}(t)-G_i) \right]
    \end{aligned} \\
    &\begin{aligned}
        \dd{\hat \rho}({\bf G}) = \mathrm{E}_W
        &\left[ \frac{-i}{\hbar}\left[ \hat{H}({\bf G}{(t)}), \hat{\rho}_c({\bf G}) \right]\dd{t} + \lambda\mathcal{D}[\hat{A}]\hat{\rho}_c({\bf G})\dd{t} + \sqrt{\lambda}\dd{W}\{ \hat{A}-\sev{\hat A}, \hat\rho({\bf G}) \} \right. + \\
        &~~\hat{\rho}_c\sum_{k=1}^n \delta^{'}(G_k^{}(t)-G_k)\left( b_k\sev{\hat A}\dd{t} + \frac{b_k\dd{W}}{\sqrt{4\lambda}} + \sum_j M_{kj}G_j{(t)}\dd{t} \right)\prod_{i\neq k}\delta(G_i^{}(t)-G_i)\\
        &~~+\hat\rho_c\sum_{k=1}^n\frac{1}{2}\delta^{''}(G_k^{}(t)-G_k)\frac{b_k^2}{4\lambda}\dd{t} \prod_{i\neq k}\delta(G_i^{}(t)-G_i) + \\
        &~~\sqrt{\lambda}\{ \hat{A} - \sev{\hat A}, \hat\rho_c \}\sum_{k=1}^n \delta^{'}(G_k^{}(t)-G_k)\frac{b_k}{\sqrt{4\lambda}}\prod_{j\neq k}\delta(G_j^{}(t)-G_j)\dd{t}+\\
        &~~\left.\hat\rho_c\sum_k\sum_{j\neq k}\delta^{'}(G_k^{}(t)-G_k)\delta^{'}(G_j^{}(t)-G_j)\frac{b_kb_j \dd{t}}{4\lambda}\prod_{i\neq k,j}^n \delta(G_i^{}(t)-G_i)\right]
    \end{aligned}
\end{align}
The stochastic terms can be neglected using $\mathrm{E}_W[dW] = 0$:
\begin{align}
    \begin{aligned}
        \dd{\hat \rho}({\bf G}) = \mathrm{E}_W&\left[ \frac{-i}{\hbar}\left[ \hat{H}({\bf G}{(t)}), \hat{\rho}_c({\bf G}) \right]\dd{t} + \lambda\mathcal{D}[\hat{A}]\hat{\rho}_c({\bf G})\dd{t}\right. + \\
        &~~\sum_{k=1}^n \delta^{'}(G_k^{}(t)-G_k)\frac{1}{2}\left\{ b_k\hat{A}\dd{t} + \sum_j M_{kj}G_j{(t)}\dd{t}, \hat{\rho}_c \right\}\prod_{i\neq k}\delta(G_i^{}(t)-G_i)\\
        &~~+\hat\rho_c\sum_{k=1}^n\frac{1}{2}\delta^{''}(G_k^{}(t)-G_k)\frac{b_k^2}{4\lambda}\dd{t} \prod_{i\neq k}\delta(G_i^{}(t)-G_i) + \\
        &~~\left.\hat\rho_c\sum_k\sum_{j\neq k}\delta^{'}(G_k^{}(t)-G_k)\delta^{'}(G_j^{}(t)-G_j)\frac{b_kb_j \dd{t}}{4\lambda}\prod_{i\neq k,j}^n \delta(G_i^{}(t)-G_i)\right]
    \end{aligned}
\end{align}

The final step is to use the following properties of the delta function \cite{Wiseman2009}:
\begin{align}
    &\frac{\partial}{\partial_{G'}}\delta(G'-G) = -\frac{\partial}{\partial_{G}}\delta(G'-G), \\
    &\left[ \frac{\partial}{\partial_{G'}}\delta(G'-G) \right] f(G') = \frac{\partial}{\partial_{G'}}\left[ \delta(G'-G) f(G) \right].
\end{align}
And the master equation derived can be rearranged like:
\begin{equation}
    \begin{aligned}
        \partial_t{\hat \rho}({\bf G}) = &\frac{-i}{\hbar}\left[ \hat{H}({\bf G}), \hat{\rho}({\bf G}) \right] + \lambda\mathcal{D}[\hat{A}]\hat{\rho}({\bf G}) - \frac{1}{2}\sum_{k=1}^n \partial_{G_k}\left\{ b_k\hat{A} + \sum_j M_{kj}G_j, \hat{\rho}({\bf G}) \right\}+\sum_k\sum_{j}\frac{b_kb_j}{8\lambda}\frac{\partial^2}{\partial_{G_k}\partial_{G_j}}\hat{\rho}({\bf G}).
    \end{aligned}
\end{equation}
Note that we collected the second derivative terms using the property of the cross-second derivative being symmetric.
This equation can also be written in a compact (vectorized) form to match Eq.~\eqref{eq:QFPME_G}:
\begin{equation}
    \partial_t{\hat \rho}({\bf G}) = \frac{-i}{\hbar}\left[ \hat{H}({\bf G}), \hat{\rho}({\bf G}) \right] + \lambda\mathcal{D}[\hat{A}]\hat{\rho}({\bf G}) - \nabla_{\bf G} \cdot \left[ \frac{1}{2} \left\{ {\bf b}\hat{A} + M{\bf G}, \hat\rho({\bf G}) \right\} \right] + \frac{1}{2}\nabla_{\bf G} \cdot \left( \frac{{\bf b}{\bf b}^T}{4\lambda} \nabla_{\bf G} \hat\rho({\bf G}) \right).
\end{equation}
The quantity ${\bf b}{\bf b}^T / 4\lambda$ represents the diffusion tensor of the ${\bf G}$-space Brownian motion.

\section{Cooling a quantum harmonic oscillator}\label{sec:appendix_cooling}
In this section, {we} describe the equations of motion used to derive the results in Sec.~\ref{sec:harmonic-oscillator}. All results were derived by solving for the asymptotic solution for the system of coupled differential equations that describe the system's energy and all other relevant expected values.

\vspace{0.5cm}
{\large \textbf{Two-layers of filtering}}

The asymptotic ensemble energy for the system that evolves under the protocol defined by Eq.~\eqref{eq:H1} is calculated by solving for the steady state of the following system:
\begin{equation}
    \partial_t \begin{pmatrix}
        R_1 \\ R_2 \\ R_3 \\ R_4
    \end{pmatrix} = 
    \begin{pmatrix}
        0 & -\Omega & 0 & 0 \\
        2\gamma & -(\Omega+\gamma) & -\Omega & -\omega \\
        0 & 2\gamma & -2(\Omega+\gamma) & 0 \\
        0 & \omega & 0 & -(\Omega+\gamma)
    \end{pmatrix}\begin{pmatrix}
        R_1 \\ R_2 \\ R_3 \\ R_4
    \end{pmatrix} + 
    \begin{pmatrix}
        \lambda \\ 0 \\ \frac{\gamma^2}{2\lambda} \\ 0
    \end{pmatrix}
\end{equation}

Each element of the vector $\vec{R} = (R_1,\dotsc,R_4)^T$ is given by a expected value:
\begin{equation}
    \begin{pmatrix}
        R_1 \\ R_2 \\ R_3 \\ R_4
    \end{pmatrix} = 
    \begin{pmatrix}
        \sev{\hat{H}_2}/\hbar\omega \\ \sev{(\hat{x}-D_{x,2})(D_{x,1}-D_{x,2})} + \sev{(\hat{p}-D_{p,2})(D_{p,1}-D_{p,2})} \\ \sev{(D_{x,1}-D_{x,2})^2} + \sev{(D_{p,1}-D_{p,2})^2} \\ \sev{(\hat{x}-D_{x,2})(D_{p,1}-D_{p,2})} - \sev{(\hat{p}-D_{p,2})(D_{x,1}-D_{x,2})}
    \end{pmatrix}.
\end{equation}

One can solve for $\sev{\hat{H}_2}_\infty$ by setting $\partial_t\vec{R} = 0$ to get Eq.~\eqref{eq:H2-infty}. Equation~\eqref{eq:ode_H2} can be calculated by taking higher order derivatives of this system of equations to isolate the equation for the internal energy.

\vspace{0.5cm}
{\large\textbf{Three-layers of filtering}}

The asymptotic energy in Eq.~\eqref{eq:H3-infty} can be calculated by setting $\partial_t \vec{S} = 0$, in the following system of equations:
\begin{equation}
\begin{aligned}
    \partial_t \vec{S} = 
    &\begin{pmatrix}
        0 & -\Omega & 0 & 0 & 0 & 0 & 0 & 0 & 0 \\
        0 & -\Omega & \omega & -\Omega & \Omega & 0 & 0 & 0 & 0 \\
        0 & -\omega & -\Omega & 0 & 0 & \Omega & 0 & 0 & 0 \\
        0 & 0 & 0 & -2\Omega & 0 &0 & 2\Omega &0 &0 \\
        2\gamma & -\gamma & 0 & 0 & -(\gamma+\Omega) & \omega & -\Omega &0 & 0 \\
        0 & 0 &-\gamma & 0 &-\omega & -(\gamma+\Omega) & 0 &-\Omega & 0 \\
        0 & \gamma & 0 &-\gamma & 0 & 0 &-(2\Omega+\gamma) & 0 &\Omega \\
        0 & 0 &-\gamma & 0 & 0 & 0 & 0 &-(2\Omega+\gamma) & 0 \\
        0 & 0 & 0 & 0 &2\gamma & 0 &-2\gamma & 0 &-2(\Omega+\gamma)
    \end{pmatrix}\vec{S}
    + &\begin{pmatrix}
        \lambda \\ 0 \\ 0 \\ 0 \\ 0 \\ 0 \\ 0 \\ 0 \\ \frac{\gamma^2}{2\lambda}
    \end{pmatrix}
\end{aligned}
\end{equation}

\begin{equation}
    \begin{pmatrix}
        S_1 \\ S_2 \\ S_3 \\ S_4 \\ S_5 \\ S_6 \\ S_7 \\ S_8 \\ S_9
    \end{pmatrix} = 
    \begin{pmatrix}
        \sev{\hat{H}_3}/\hbar\omega \\ \sev{(\hat{x}-D_{x,3})(D_{x,2}-D_{x,3})} + \sev{(\hat{p}-D_{p,3})(D_{p,2}-D_{p,3})} \\ \sev{(\hat{p}-D_{p,3})(D_{x,2}-D_{x,3})} - \sev{(\hat{x}-D_{x,3})(D_{p,2}-D_{p,3})} \\ \sev{(D_{x,2}-D_{x,3})^2} + \sev{(D_{p,2}-D_{p,3})^2} \\ \sev{(\hat{x}-D_{x,3})(D_{x,1}-D_{x,2})} + \sev{(\hat{p}-D_{p,3})(D_{p,1}-D_{p,2})} \\ \sev{(\hat{p}-D_{p,3})(D_{x,1}-D_{x,2})} - \sev{(\hat{x}-D_{x,3})(D_{p,1}-D_{p,2})} \\ \sev{(D_{x,1}-D_{x,2})(D_{x,2}-D_{x,3})} + \sev{(D_{p,1}-D_{p,2})(D_{p,2}-D_{p,3})} \\ \sev{(D_{p,2}-D_{p,3})(D_{x,1}-D_{x,2})} - \sev{(D_{x,2}-D_{x,3})(D_{p,1}-D_{p,2})} \\ \sev{(D_{x,1}-D_{x,2})^2} + \sev{(D_{p,1}-D_{p,2})^2}
    \end{pmatrix}.
\end{equation}

\vspace{0.5cm}
{\large\textbf{Band-pass filtering}}

The asymptotic energy of the band-pass filter system in Eq.~\eqref{eq:band_pass_energy} can be calculated by solving:
\begin{equation}
    \label{eq:appendix_system_HBP}
    \partial_t \begin{pmatrix}
        T_1 \\ T_2 \\ T_3 \\ T_4 \\ T_5 \\ T_6 \\ T_7 \\ T_8 \\ T_9
    \end{pmatrix} = 
    \begin{pmatrix}
        -2\gamma & \Omega & 0 & 0 & 0 & 0 & 0 & 0 & 0 \\
        0 & -2\gamma & -\omega & \Omega & \Omega & 0 & 0 & 0 & 0 \\
        0 & \omega & -2\gamma & 0 & 0 & \Omega & 0 & 0 & 0 \\
        0 & 0 & 0 & -2\gamma & 0 & 0 & 2\Omega & 0 & 0 \\
        2\gamma & -\Omega & 0 & 0 &-\gamma &-\omega &\Omega & 0 & 0 \\
        0 & 0 & -\Omega & 0 &\omega & -\gamma & 0 &\Omega & 0 \\
        0 & \gamma & 0 & -\Omega & 0 & 0 &-\gamma & 0 & \Omega \\
        0 & 0 & -\gamma & 0 & 0 & 0 & 0 &-\gamma & 0 \\
        0 & 0 & 0 & 0 & 2\gamma & 0 &-2\Omega & 0 & 0
    \end{pmatrix}\vec{T} + 
    \begin{pmatrix}
        \frac{1}{2}\left( \frac{\lambda}{\gamma} + \frac{\gamma}{4\lambda} \right) \\ 0 \\ 0 \\ 0 \\ -\frac{\gamma^2}{2\lambda} \\ 0 \\ 0 \\ 0 \\ \frac{\gamma^2}{2\lambda}
    \end{pmatrix}
\end{equation}

\begin{equation}
    \begin{pmatrix}
        T_1 \\ T_2 \\ T_3 \\ T_4 \\ T_5 \\ T_6 \\ T_7 \\ T_8 \\ T_9
    \end{pmatrix} = 
    \begin{pmatrix}
        \sev{\hat{H}({\bf E_1})}/\hbar\omega \\ \sev{(\hat{x}-E_{x,1})E_{x,2}} + \sev{(\hat{p}-E_{p,1})E_{p,2}} \\ \sev{(\hat{x}-E_{x,1})E_{p,2}} - \sev{(\hat{p}-E_{p,1})E_{x,2}} \\ \sev{E_{x,2}^2+E_{p,2}^2} \\ \sev{(\hat{x}-E_{x,1})E_{x,1}} + \sev{(\hat{p}-E_{p,1})E_{p,1}} \\ \sev{(\hat{x}-E_{x,1})E_{p,1}} - \sev{(\hat{p}-E_{p,1})E_{x,1}} \\ \sev{E_{x,1}E_{x,2} + E_{p,1}E_{p,2}} \\ \sev{E_{x,2}E_{p,1} - E_{p,2}E_{x,1}} \\ \sev{E_{x,1}^2 + E_{p,1}^2}
    \end{pmatrix}.
\end{equation}

\end{document}